\documentstyle[12pt]{article}
\begin{document}
\topmargin = 0.0 truecm
\voffset=-1.0 truecm
\hoffset=0.0 truecm
\begin{center}

{\Large Strangeness Production }
\medskip
{\Large in Chemically Non-Equilibrated Parton Plasma}

\vskip 0.8cm

{\large P\'eter L\'evai$^a$\footnote{ The talk was presented
by P. L\'evai on S'95 Conference, Tucson, Jan. 3-7 1995.
E-mail:plevai@sunserv.kfki.hu} and Xin-Nian Wang$^b$}
\vskip 0.6cm

{ \it $^a$Research Institute for Particle and Nuclear Physics }

{ \it 1525 Budapest 114. POB. 49, Hungary }
\medskip

{ \it $^b$Nuclear Science Division, MS 70A-3307 }

{ \it Lawrence Berkeley Laboratory, Berkeley, CA 94720}

\end{center}
\vskip 0.5 cm

\noindent {\small ABSTRACT:  Strangeness production was investigated
during the equilibration of a gluon dominated parton plasma
produced at RHIC and LHC energies. The time evolution of parton densities
are followed by a set of rate equations in a 1-dimensional expanding
system. The strangeness production will depend on the initial chemical
equilibration level and in our case the parton densities will remain
far from the full equilibrium. We investigate the influence of gluon
fragmentation on final strangeness content.  }

\vspace{0.5cm}

\section*{Introduction}

Strangeness production was widely investigated during last decade as
one of the possible quark-gluon plasma signatures \cite{KMR}.
To refresh our knowledge about the time dependence of
strangeness content in case of parton gas, let us consider
ultrarelativistic heavy ion collisions in the framework of pQCD-inspired
parton interaction models \cite{HIJING,PCM,DTUNUC}.
In this framework, hard or semihard
scatterings among partons dominate the reaction dynamics. They can liberate
partons from the individual nucleons, thus producing large
amount of transverse energy in the central region \cite{JBAMKLL},
and drive the initially produced parton system toward
equilibrium \cite{PCM,SHUR93,BDMTW}.

Roughly speaking, ultrarelativistic heavy-ion collisions in
a partonic picture can be divided into three stages: (1)
During the early stage, hard or semihard parton scatterings,
which happen on a time scale of about $0.2$ fm/$c$, produce a hot
and undersaturated parton gas. This parton gas is dominated by gluons and
its quark content is far below the chemical equilibrium value.
Interference and parton fusion also leads to the depletion
of small $x$ partons in the effective parton distributions
inside a nucleus, reducing the initial parton production \cite{BEW}.
(2) After two beams of partons pass through each other, the produced
parton gas in the central rapidity region starts its evolution
toward (kinetic) thermalization mainly through elastic
scatterings and expansion. The kinematic separation
of partons through free-streaming gives an estimate of the time scale
$\tau_{\rm iso}\sim 0.5 - 0.7 \  fm/c$ \cite{BDMTW,KEXW94a},
when local isotropy in momentum distributions is reached.
(3) Further evolution of the parton gas toward
a fully (chemically) equilibrated parton plasma is
dictated by the parton proliferation through induced
radiation and gluon fusion. This chemical equilibration can be
described by a set of rate equations for gluons and quarks with
different flavours.
Due to the consumption of
energy by the additional parton production, the effective temperature
of the parton plasma cools down considerably faster than the ideal
Bjorken's scaling solution. Therefore, the life time of the plasma
is reduced before the temperature drops below the
QCD phase transition temperature \cite{BDMTW}.

Strangeness production could also be divided into the above three
different contributions in the history of the evolution of the parton system.
However, here we will skip the very complicated description of steps (1) and
(2) and simply we consider the result of the HIJING Monte Carlo code for
these early stages. We will concentrate on strangeness production in stage
(3), investigating the strangeness evolution and saturation. We
describe the equilibration of the initially produced hot and undersaturated
parton
gas following Ref.~\cite{BDMTW}. For gluons we apply the result of an
improved perturbative QCD analysis of Landau-Pomeranchuk-Migdal
effect (see Ref. \cite{GWLPM} and Ref. \cite{OCHARM}).

In our analysis we include one more step of strangeness production, namely
the gluon fragmentation. We assume a two step hadronization process
at a certain critical temperature:
(a) gluons decay into quark-antiquark pairs forming quark-matter from the
parton gas; (b) the quarks and antiquarks will be redistributed forming
colorless hadrons.
For simplicity, we will use the well-known string breaking factors for
estimating the different flavour yields in the gluon decay reaction
($g \longrightarrow q_i + {\overline q}_i$). However more work will be
needed to find out the medium-dependent values of these factors.
The hadron formation by quark-antiquark redistribution is beyond of
the scope of our recent investigation. However, the strange/non-strange
quark-pair ratio will not change in this type of hadronization. One can find
a detailed description of such a hadronization mechanism in Ref.
\cite{ALCOR}.

We will study the strangeness production during the parton
evolution. We also discuss the influence of the uncertainties in
the initial parton density on the final saturation level of different
flavours. We investigate the extra strangeness yield of the gluon
fragmentation and we compare the change of the strange/non-strange
quark ratio in a chemically non-equilibrated parton system to
a fully equilibrated one.

\section*{Initial conditions: \ hot \ and \ undersaturated \ \ \
 parton gas}

The initial conditions for the time evolution of
thermalized but chemically non-equilibrated
parton gas can be obtained from one of the microscopical models.
Currently there are many models for incorporating hard and semi-hard
processes in hadronic and nuclear collisions \cite{HIJING,PCM,DTUNUC}.
Here we will use the result of the HIJING Monte Carlo model \cite{HIJING}
to estimate the initial parton production at time, $\tau_{\rm iso}$.

Since we are primarily interested in the chemical equilibration
of the parton gas which has already reached local isotropy in momentum
space, we shall assume that a parton distribution
can be approximated by thermal phase space distribution with
non-equilibrium fugacity $\lambda $:
\begin{equation}
f(k;T,\lambda)  = \lambda \left( e^{u\cdot k /T} \pm
\lambda \right)^{-1}, \label{eq1}
\end{equation}
where $u^{\mu}$ is the
four-velocity of the local comoving reference frame.  When a
parton fugacity $\lambda$ is much less than unity as
may happen during the early evolution of the parton system,
we can neglect the quantum corrections in Eq.~(\ref{eq1}) and
write the momentum distributions in the factorized form,
\begin{equation}
\label{19}
f(k;T,\lambda )=\lambda \left (e^{ u\cdot k /T}\pm 1\right)^{-1}.
\label{eq2}
\end{equation}
Using this distribution in the comoving
frame for a multi-component parton gas,
one can obtain the total parton density, energy density and pressure as
\begin{eqnarray}
n &=& n_g + \sum_i (n_{q_i} + n_{{\overline q}_i}) =
(\lambda_g a_1 + \sum_i \lambda_i b_1(x_i) ) \cdot T^3 \label{eq3} \\
\quad \varepsilon &=& \varepsilon_g +
\sum_i (\varepsilon_{q_i} + \varepsilon_{{\overline q}_i}) =
(\lambda_g a_2 + \sum_i \lambda_i b_2(x_i) ) \cdot T^4 \label{eq4} \\
\quad P &=& P_g + \sum_i (P_{q_i} + P_{{\overline q}_i} ) =
{1\over 3} (\lambda_g a_3  +
 \sum_i \lambda_i b_2(x_i) b_3(x_i) ) \cdot T^4 . \label{eq5}
\end{eqnarray}
Here $a_1=16\zeta (3)/\pi^2\approx 1.95$ and
$a_2=a_3=8\pi^2/15\approx 5.26$
for the Bose distribution of gluons.
For fermions ($i=u, d, s, $)
we introduce the parameter $x_i = m_i/T$.
For massless light quark-antiquark pairs one obtains
$b_1(0)=2 \cdot 9\zeta (3)/2 \pi^2\approx 1.10$,
$b_2(0)=2 \cdot 7\pi^2/40 \approx 3.44$ and $b_3(0)=1$. For massive
strange quark-antiquark pairs one can determine these factors exactly (see in
Appendix).
Here we investigate baryon symmetric systems, which means
$\lambda_u=\lambda_{\bar u}$,
$\lambda_d=\lambda_{\bar d}$ and $\lambda_s=\lambda_{\bar s}$.
Since boost invariance
has been demonstrated to be a good approximation for
the initially produced partons \cite{KEXW94a}, we can
estimate the initial parton fugacities, $\lambda_{g}^0, \lambda_{i}^0$
and temperature $T_0$  from
\begin{equation}
n_0 = \frac{1}{\pi R^2_{A} \tau_{\rm iso}} \frac{dN}{dy}\; ,
\quad \varepsilon_0 = n_0 \frac{4}{\pi}\langle k_T\rangle, \label{eq6}
\end{equation}
where $\langle k_T\rangle$ is the average transverse momentum.
{}From HIJING one can obtain the initial values of the
different gluon and quark
fugacities. For different flavours
the initial ratio is $u:d:s \approx 2:2:1$, approximately.

 Table 1. shows these relevant quantities at the moment $\tau_{\rm iso}$,
for Au + Au collisions at RHIC and LHC energies. One can observe
that the initial parton gas is rather hot reflecting the large
average transverse momentum. However, the parton gas is
undersaturated as compared to the ideal gas at the same temperature. The
gas is also  dominated by gluons with its quark content far below
the chemical equilibrium value. We should emphasize that the initial
conditions listed here result from HIJING
calculation of parton production through semihard scatterings.
Soft partons, {\em e.g.}, due to parton production from the color
field \cite{KEMG}, are not included.

\begin{center}
\begin{tabular}{||l||l|l||}
\hline  {\bf Au+Au} &  {\bf RHIC} & {\bf LHC}  \\
\hline
\hline
 $\tau_{iso}$(fm/$c$)   & \ 0.7 & \ 0.5   \\
\hline
 $\varepsilon_0$ (GeV/fm$^3$)& \ 3.2 & 40   \\
\hline
 $n_0$(fm$^{-3}$)     & \ 2.15 & 18   \\
\hline
 $\langle k_\perp \rangle $ (GeV)& \ 1.17 & \ 1.76 \\
\hline
 $T_0$ (GeV)          & \ 0.55 & \ 0.82 \\
\hline
 $\lambda_g^0$        & \ 0.05 & \ 0.124 \\
\hline
 $\lambda_u^0$        & \ 0.004 & \ 0.01 \\
\hline
 $\lambda_d^0$        & \ 0.004 & \ 0.01 \\
\hline
 $\lambda_s^0$        & \ 0.002 & \ 0.005 \\
\hline
\end{tabular}
\end{center}
\vskip 0.5cm

\noindent {\bf Table 1:}
Values of the relevant parameters characterizing the parton plasma at the
moment $\tau_{iso}$, when local isotropy of the momentum distribution is first
reached.

\vskip 0.3cm

\section*{Master rate equations}

In general, chemical reactions among partons can be quite complicated
because of the possibility of initial and final-state gluon radiations.
Interference effects due to multiple scattering inside a dense medium,
{\em i.e.}, LPM  suppression of soft gluon radiation has to be taken
into account. One lesson one has learned from LPM effect \cite{GWLPM}
is that the radiation between two successive scatterings is the
sum, {\em on the amplitude level}, of both the initial state
radiation from the first scattering and the final state radiation
from the second one. Since the off-shell parton is space-like
in the first amplitude and time-like in the second, the picture
of a time-like parton propagating inside a medium in the parton
cascades simulations \cite{PCM} shall break down. Instead,
we shall here consider both initial and final state radiations
together associated with a single scattering (to the leading order,
a single additional gluon is radiated, such as $gg\to ggg$),
in which we can include LPM effect by a radiation suppression
factor. The analysis of QCD LPM effect in Ref.~\cite{GWLPM}
has been done for a fast parton traveling inside a parton
plasma. We will use the results for radiation off thermal
partons who average energy is about $T$, since we expect
the same physics to happen.

In order to permit the approach to chemical equilibrium, the reverse
process, {\em i.e.}, gluon absorption, has to be included as well, which is
easily achieved making  use of detailed balance.
We consider only the dominant process $gg\to ggg$.
Radiative processes involving quarks have substantially smaller
cross sections in pQCD, and quarks are considerably less
abundant than gluons in the initial phase of the chemical evolution of
the parton gas.  Here we are interested in understanding the basic
mechanisms underlying the formation of a chemically equilibrated
quark-gluon plasma, and the essential time-scales.  We hence restrict
our considerations to the dominant reaction mechanisms for the
equilibration of each parton flavor.  These are just four
basic processes \cite{MSM86}:
\begin{equation}
gg \leftrightarrow ggg, \quad gg\leftrightarrow
q_i\overline{q_i}.\label{eq7}
\end{equation}
Other scattering processes ensure the maintenance of thermal
equilibrium $(gg\leftrightarrow gg, \; gq_i \leftrightarrow gq_i$, etc.)
or yield corrections to the dominant reaction rates
$(gq_i\leftrightarrow q_igg$, etc.).

Restricting to the reactions Eq.~(\ref{eq7}) and assuming
that elastic parton scattering is sufficiently rapid to maintain
local thermal equilibrium, the evolution of the parton densities
is governed by the master equations \cite{BDMTW}:
\begin{eqnarray}
\partial_{\mu}(n_gu^{\mu}) &= &
 \frac{1}{2}\sigma_3 n_g^2 \left( 1-\frac{n_g}{\tilde n_g}\right)
 -\sum_i 2 \frac{1}{2}\sigma_2^i n_g^2 \left( 1 - \frac{n_{q_i}^2 \tilde n_g^2}
 {\tilde n_{q_i}^2 n_g^2}\right), \label{eq8}\\
 \partial_{\mu} (n_{q_i}u^{\mu}) &= &\partial_{\mu} (n_{\bar{q}_i} u^{\mu})
 =\frac{1}{2}\sigma_2^i n_g^2 \left( 1 - \frac{n_{q_i}^2 \tilde n_g^2}
 {\tilde n_{q_i}^2 n_g^2}\right), \label{eq9}
\end{eqnarray}
where ${\tilde n_g}\equiv n_g/\lambda_g$ and
${\tilde n_{q_i}}\equiv n_{q_i}/\lambda_i$
denote the densities
with unit fugacities, $\lambda_g=\lambda_{q_i}=1$, $\sigma_3$ and $\sigma_2^i$
are thermally averaged, velocity weighted cross sections,
\begin{equation}
\sigma_3 = \langle\sigma(gg\to ggg)v\rangle, \quad \sigma_2^i =
\langle \sigma (gg\to q_i{\overline{q}}_i )v\rangle. \label{eq10}
\end{equation}
In Eq.~(\ref{eq9}) we neglected the exchange between quark flavours, because
generally the gluonic channel is dominant.
We have also assumed detailed balance and a baryon symmetric
matter, $n_{q_i}=n_{{\bar q}_i}$. If we neglect effects of viscosity
due to elastic scattering \cite{KEMG,VISC}, we then have the
hydrodynamic equation
\begin{equation}
\partial_{\mu} (\varepsilon u^{\mu}) + P\;\partial_{\mu} u^{\mu} = 0,
\label{eq11}
\end{equation}
which determines the evolution of the energy density.

For a time scale $\tau\ll R_A$, we can neglect the transverse
expansion and consider the expansion of the parton plasma
purely longitudinal, which leads to the Bjorken's scaling
solution \cite{BJOR} of the hydrodynamic equation:
\begin{equation}
{d\varepsilon\over d\tau} + {\varepsilon+P\over\tau} = 0. \label{eq12}
\end{equation}

Evaluating this  ultrarelativistic equation of motion
by means of Eqs. (\ref{eq4})-(\ref{eq5}) one obtains
\begin{equation}
 \left[ {{\dot{\lambda}_g} \over \lambda_g} + 4 {{\dot T} \over T} +
{1 \over \tau} {4\over 3} \right] \varepsilon_g +
\sum_i \left[ {{\dot{\lambda}_i} \over \lambda_i} + 4   {{\dot T} \over T}
 D_\varepsilon(x_i) +
{1 \over \tau} \left( 1 + {1\over 3} b_3(x_i) \right) \right] \varepsilon_i
= 0   \label{eq13}
\end{equation}
where $ \varepsilon_i = \varepsilon_{q_i} + \varepsilon_{{\overline q}_i}$ and
$D_\varepsilon(0)= b_3(0)=1$ for massless quark-antiquark pairs,
$D_\varepsilon(x_s)$, \ $b_3(x_s)$ for massive strange quark can be found in
Appendix. Rewriting the rate equations Eqs.~(\ref{eq7})-(\ref{eq8})
in terms of time evolution of the
parameters $T(\tau)$, $\lambda_g(\tau)$ and $\lambda_i(\tau)$, one obtains
\begin{eqnarray}
\frac{\dot\lambda_g}{\lambda_g} + 3\frac{\dot T}{T} + \frac{1}{\tau} &=
&R_3 (1-\lambda_g)-
2 \sum_i R_2^i \left(1- \frac{\lambda_i^2}{\lambda_g^2} \right)
        \label{eq14} \\
\frac{\dot\lambda_i}{\lambda_i} + 3\frac{\dot T}{T} D_n(x_i)
+ \frac{1}{\tau} &=
&R_2^i {n_g\over n_{q_i}} \left( 1 - \frac{\lambda_i^2}{\lambda_g^2} \right)
\label{eq15}
\end{eqnarray}
where the density weighted reaction rates $R_3$ and $R_2^i$ are defined as
\begin{equation}
R_3 = \textstyle{{1\over 2}} \sigma_3 n_g, \quad
R_2^i = \textstyle{{1\over 2}} \sigma_2^i n_g.  \label{eq16}
\end{equation}
Notice that for a fully equilibrated system ($\lambda_g=\lambda_i=1$),
Eq. (\ref{eq13}) corresponds to the Bjorken solution,
$T(\tau)/T_0=(\tau_0/\tau)^{1/3}$. In Eq. (\ref{eq15}) $D_n(0) = 1$ and
for $D_n(x_s)$ see Appendix.

\section*{Parton equilibration rates}

To take into account of the LPM effect in the calculation of
the reaction rate $R_3$ for $gg\rightarrow ggg$,
we simply impose the LPM suppression of
the gluon radiation whose effective formation time $\tau_{\rm QCD}$
is much longer than the mean-free-path $\lambda_f$ of multiple
scatterings to each $gg\rightarrow ggg$ process.
In the mean time, the LPM effect also regularizes
the infrared divergency associated with QCD radiation. However,
$\sigma_3$ still contains infrared singularities in the gluon
propagators. For an equilibrium system one can in principle apply
the resummation technique developed by Braaten and Pisarski \cite{BP90}
to regularize the electric part of the propagators, though the
magnetic sector still has to be determined by an unknown magnetic
screening mass which can only be calculated nonperturbatively \cite{TBBM93}
up to now. Since we are dealing with a non-equilibrium system,
Braaten and Pisarski's resummation may not be well defined. As an
approximation, we will use the Debye screening mass \cite{BMW92},
\begin{equation}
\mu_D^2 = {6g^2\over \pi^2} \int_0^{\infty} kf(k) dk
=4\pi\alpha_s T^2\lambda_g, \label{eq17}
\end{equation}
to regularize all singularities in both the scattering cross
sections and the radiation amplitude.

The mean-free-path for elastic scatterings is
\begin{equation}
  \lambda_f^{-1}\equiv n_g\int_0^{s/4}dq_{\perp}^2
    \frac{d\sigma_{\rm el}^{gg}}{dq_{\perp}^2}
  = n_g\int_0^{s/4}dq_{\perp}^2
    \frac{9}{4}\frac{2\pi\alpha_s^2}{(q_{\perp}^2+\mu_D^2)^2}
  = \frac{9}{8}a_1\alpha_s T\frac{1}{1+8\pi\alpha_s\lambda_g/9}, \label{eq18}
\end{equation}
which depends very weekly on the gluon fugacity $\lambda_g$ \cite{OCHARM}.

The modified differential cross section for $gg\rightarrow ggg$ has the form:
\begin{equation}
  \frac{d\sigma_3}{dq_{\perp}^2 dy d^2k_{\perp}}
  =\frac{d\sigma_{\rm el}^{gg}}{dq_{\perp}^2}\frac{dn_g}{dy d^2k_{\perp}}
  \theta(\lambda_f-\tau_{QCD})\theta(\sqrt{s}-k_{\perp}\cosh y), \label{eq19}
\end{equation}
where the second step-function accounts for energy conservation,
$s=18T^2$ is the average squared center-of-mass energy of two
gluons in the thermal gas and the LPM suppression factor was approximated by a
$\theta$-function, $\theta(\lambda_f-\tau_{\rm QCD})$, where
$\tau_{\rm QCD}= \cosh y / k_{\perp}$ is the effective formation time of the
gluon radiation in SU(3) ~\cite{GWLPM}.
We can complete part of these integrations:
\begin{equation}
  R_3/T=\frac{32}{3a_1}\alpha_s\lambda_g(1+8\pi\alpha_s\lambda_g/9)^2
  {\cal I}(\lambda_g), \label{eq20}
\end{equation}
where ${\cal I}(\lambda_g)$ is a function of $\lambda_g$,
\begin{eqnarray}
{\cal I}(\lambda_g)=\int_1^{\sqrt{s}\lambda_f}dx
\int_0^{s/4\mu_D^2}&dz& \frac{z}{(1+z)^2}
  \left\{ {\cosh^{-1}(\sqrt{x}) \over
  x\sqrt{[x+(1+z)x_D]^2-4x\;z\;x_D}}\right. \nonumber \\
  &+& \left. \frac{1}{s\lambda_f^2}{\cosh^{-1}(\sqrt{x}) \over
  \sqrt{[1+x(1+z)y_D]^2-4x\;z\;y_D}}\right\}, \label{eq21}
\end{eqnarray}
where $x_D=\mu_D^2\lambda_f^2$, $y_D=\mu_D^2/s$ and the integral
is evaluated numerically ~\cite{OCHARM}.

Note that in principle one should multiply the phase-space
integral by $1/3!$ to take into account of the symmetrization
of identical particles in $gg \rightarrow ggg$ as in Ref.~\cite{SHUR93}.
However, for the dominant soft radiation we consider here,
the radiated soft gluon does not overlap with the two incident
gluons in the phase-space. Thus we only multiply the cross section
by $1/2!$.

The quark equilibration rates $R_2^i$ for
$gg\rightarrow q_i{\bar q}_i$ can be also calculated, following the
approximations
of Ref.~\cite{BDMTW}. We introduce the thermal quark mass as a cutoff in the
logaritmically divergent integrals:
\begin{equation}
M_i^2 = m_{0,i}^2 + \left[ \lambda_g +\frac{1}{2} \left(
\lambda_u + \lambda_d + \lambda_s \right) \right] \frac{4 \pi}{9}
\alpha_s T^2 \ . \label{eq22}
\end{equation}
Here $m_{0,i}$ notes the bare quark mass, $m_{0,u}=m_{0,d} \approx 0 $,
$m_{0,s} = 150 \ MeV$.

The total cross section is dominated by the Compton diagrams. The
cross section is the following for the process
$gg \longrightarrow q_i {\overline q}_i$:
\begin{equation}
{ {d \sigma_i} \over {dt}} = { {\pi \alpha_s^2} \over {3 s}}
\left[ \log {{1+\chi} \over {1-\chi}} \left( 1 + \frac{1}{2} {{M_i^2}\over s}
- {{M_i^4} \over s^2} \right) + \chi \left( - \frac{7}{4} - \frac{31}{4}
\frac{M_i^2}{s} \right) \right] \ \ \label{eq23}
\end{equation}
where $\chi = \sqrt{1 - 4 M_i^2/s}$.

To obtain the total cross section approximately,
we integrate over thermal gluon distributions and we insert
the thermal gluon mass, $M_i$, and
the average thermal $\langle s \rangle = 18 T^2 $
into the cross section:
\begin{equation}
\sigma^i_2 \approx \langle \sigma_i v \rangle \approx
\frac{9}{4} \left( \frac{\zeta (2)}{\zeta (3)} \right)^2
\left. \frac{d \sigma_i}{dt}\right|{}_{M_i; \ \langle s \rangle = 18 T^2}
\label{eq24}
\end{equation}

By means of Eq. (\ref{eq24}) one can derive numerically the production rates
$R^i_2 = 1/2 \cdot \sigma^i_2 n_g$ which will be used in Eq.
(\ref{eq14})-(\ref{eq15}) in the calculation of the time evolution.

\section*{Evolution of the parton plasma}

With the parton equilibration rates which in turn depend
on the parton fugacity, we can solve the master equations
self-consistently and obtain the time evolution of the temperature
and the fugacities. The time dependence of $T$, $\lambda_g$ and $\lambda_i$
are shown in Fig. 1.a,
and Fig. 2.a, for initial conditions listed in Table I. at RHIC and
LHC energies. We find that the parton gas cools considerably
faster than predicted by Bjorken's scaling solution
($(T^*)^3\tau$ = const.) shown as dotted lines, because
the production of additional partons
approaching the chemical equilibrium state consumes an appreciable
amount of energy. The accelerated cooling, in turn, slows down the
chemical equilibration process, which is more apparent at RHIC
than at LHC energies. Therefore, the parton system can hardly
reach its equilibrium state before the effective temperature
drops below $T_c \approx 200$ MeV in a short period of time of
2-3 fm/$c$ at RHIC energy. At LHC energy, however, the parton
gas will be closer to its equilibrium and the plasma
may exist in a deconfined phase for as long as 4-5 fm/$c$.

We note that the initial conditions used here are the results
of the HIJING model calculation in which only initial direct parton
scatterings are taken into account.  Due to the fact that HIJING is
a QCD motivated phenomenological model, there are some uncertainties
related to the initial parton production, as listed in Ref.~\cite{BDMTW}.
We can estimate the effect of the uncertainties in the
initial conditions on the parton gas evolution by multiplying
the initial energy and parton number densities at RHIC and LHC energies
by a factor of 4. This will increase the initial fugacities
approximately by a factor of 4 at the same initial temperature.
With these high initial densities, the parton plasma can evolve closer to
an equilibrated gluon gas as shown in Fig. 1.b, and Fig. 2.b,
however the system is still non-equilibrated and
dominated by gluons. The light quarks and
strange quarks remain far from the full chemical equilibrium.

{\bf Thus we can conclude that perturbative parton
production and scatterings are very likely to produce
a fully equilibrated parton plasma
in ultrarelativistic heavy ion collisions at RHIC and LHC energies.
The strangeness content will be far from the equilibrium, also.}

\section*{Gluon fragmentation}

We have seen that during the parton
evolution strangeness and the other quarks
were not produced at equilibrium level. Thus the last possibility to get some
more strangeness is the gluon fragmentation before hadronization.
In an oversimplified model we can investigate the relative strangeness
enhancement or suppression in the hadronization.
We assume a two step hadronization process
at a certain critical temperature:
(a) gluons decay into quark-antiquark pairs forming quark-matter from the
parton gas; (b) the quarks and antiquarks will be redistributed forming
colorless hadrons.

For simplicity, we will use the well-known string breaking factors to
estimate the different flavour yields in the gluon decay process
($g \longrightarrow q_i + {\overline q}_i$), namely
$f_u=f_d=0.425$ and $f_s=0.15$. Generally, in a hot and under-saturated
parton gas these factors will not be valid automatically, but in lack of
correct, temperature and parton density dependent values, here we will use
the above numbers. Let us define the relative strangeness content by the
strange/non-strange flavour ratio $g_s$:
\begin{equation}
g_s = { n_s \over {n_u + n_d}}
\end{equation}

One can determine easily this ratio $g_s$ in the non-equilibrated parton gas
before and after fragmentation and we can compare its value to a fully
equilibrated quark-gluon plasma characterized by the same temperature.
Fig.1.c, d, and Fig.2.c, d, show the results of our calculation at RHIC and
LHC energies.

On Fig.1.c, full lines display the ratio $g_s$ for the previously calculated
gluon dominated non-equilibrium parton gas (smaller value, $g_s \approx 0.25$
and marked as {\it "Non-eq."}) and for the fully equilibrated quark-gluon
plasma (larger value, $g_s \approx 0.45$ and marked as {\it "Eq."}).
Gluon fragmentation, characterized by above values $f_i$,
leads to the enhancement of the light quark flavours and the suppression of
the strangeness. On Fig.1.c, after fragmentation $g_s$ will drop in both
non-equilibrium  (dotted line marked as {\it "Fr-Non-eq"} and
$g_s \approx 0.18$) and equilibrium cases (dashed line marked as
{\it "Fr-Eq"} and $g_s \approx 0.28$). Thus strangeness will be suppressed
during hadronization in any case and any time step.
Even if we consider  4 times higher initial parton fugacities (see Fig.1.d,)
then we will have very much the same result as before. Moreover, the
final $g_s$ ratio will depend very weakly on the initial condition.

On Fig.2.c, we repeated the gluon fragmentation at LHC energy.
We obtained an $\approx 25 \%$ increase in strangeness comparing to
the RHIC energy and an evolution which is slightly
closer to the equilibrium ones.
On Fig.2.d, the 4 times larger initial parton fugacities display
a similar increase in strangeness, however the strangeness ratio
will also drop during hadronization.

Our conclusions are the following:
(a) before fragmentation $g_s$ is far from its equilibrium value
and it depends very strongly on the initial parton fugacities;
(b) the above simple gluon fragmentation will decrease this $g_s$ factor,
suppressing the strangeness in a quark matter;
(c) after fragmentation  in equilibrium case $g_s \approx 0.3$,
in non-equilibrium case
$g_s \approx 0.2$ at RHIC energy and $g_s \approx 0.25$ at LHC energy.

This factor $g_s$ is important,  because after gluon
fragmentation the above flavour ratio will not change during the
quark redistribution and it will determine the final strange/non-strange
hadron ratio.
The hadron formation via quark and antiquark redistribution is beyond of
the scope of our recent investigation.  One can find
a detailed description of such a hadronization mechanism in Ref.
\cite{ALCOR} where at 200 GeV/n bombarding energy (CERN SPS)
it was extracted  $g_s =0.16$ for nucleon-nucleon and $g_s=0.255$ for
S+S collision.

\section*{Conclusions}

In this paper, we have calculated strangeness production
in an equilibrating parton plasma, taking into account
the evolution of the effective temperature and parton
fugacities according to the solution of a set of rate
equations. In the evaluation of the interaction rate
$R_3$ for induced gluon radiation, a color dependent
effective formation time was used which reduces the
gluon equilibration rate through LPM suppression of
soft gluons.

We found that the thermal contribution during
the parton equilibration
with the current estimate of the initial parton density from
HIJING Monte Carlo simulation are larger than the
initial direct strangeness production, but the reached saturation level
remains far from the full equilibrium at RHIC and LHC energies.
The strangeness production depends on the initial
condition of the parton evolution. If uncertainties in
the initial parton production can increase the initial
parton density, {\em e.g.}, by a factor of 4,
the secondary strangeness production will be larger,
but even in this case it will not reach the total chemical
equilibrium.

However, during hadronization the yield of light quarks
can be much larger as the strange one. Thus the hadronization process
may lead to a strangeness suppression. This conclusion  very much
depends on the fragmentation ratios.
The strange/non-strange quark ratio will be also far from its equilibrium
value during the partonic evolution. Moreover it could drop during
hadronization and the non-equilibrium characteristic will be conserved.


\section*{Appendix}

For massive partons we have used the following expressions:

\begin{eqnarray}
 b_1(x_i)&=&2 {d_i \over {2 \pi^2}} \cdot x_i^3
  \sum_{n=1}^\infty (-1)^{n+1} {1 \over {n x_i}} K_2(n x_i)  \label{a1} \\
 b_2(x_i)&=&2 {d_i \over {2 \pi^2}} \cdot x_i^4
  \sum_{n=1}^\infty (-1)^{n+1} \left[
  {3 \over {(n x_i)^2}} K_2(n x_i) +
  {1 \over {(n x_i)  }} K_1(n x_i) \right] \label{a2} \\
 b_3(x_i) &=&
 { {  \sum_{n=1}^\infty (-1)^{n+1} {1 \over {(n x_i)^2}} K_2(n x_i) }
  \over  { \sum_{n=1}^\infty (-1)^{n+1} \left[
    {1 \over {(n x_i)^2}} K_2(n x_i) +
    {1 \over 3} {1 \over {(n x_i)  }} K_1(n x_i)  \right]  }} \label{a3}
\end{eqnarray}

\begin{eqnarray}
 D_n(x_i) &=&
 { {  \sum_{n=1}^\infty (-1)^{n+1} \left[ {1 \over {(n x_i)}} K_2(n x_i)
      + {1 \over 3} K_1(n x_i) \right] }
  \over  { \sum_{n=1}^\infty (-1)^{n+1} {1 \over {(n x_i)}}
       K_2(n x_i)  }} \label{a4} \\
 D_\varepsilon(x_i) &=&
 { {  \sum_{n=1}^\infty (-1)^{n+1} \left[
{1 \over {(n x_i)^2}} K_2(n x_i) +
{5 \over{12}} {1 \over {(n x_i)}} K_1(n x_i) +
{1 \over{12}} K_0(n x_i) \right] }
  \over  { \sum_{n=1}^\infty (-1)^{n+1} \left[
  {1 \over {(n x_i)^2}} K_2(n x_i) +
  {1 \over 3} {1 \over {(n x_i)}}   K_1(n x_i) \right]  }} \ \ \ \ \ \label{a5}
\end{eqnarray}

with $x_i=m_i/T$ and $d_i=6$. The factor $(-1)^{n+1}$ is connected
to fermions.

\section*{Acknowledgments}

P. L. would like to thank  T. S. Bir\'o, B. Kaempfer and J. Zim\'anyi
for helpful  discussions.
P. L. was supported by the Hungarian Science Fund, OTKA No. T014213.
X.-N. W. was supported by the Director, Office of Energy
Research, Division of Nuclear Physics of the Office of High
Energy and Nuclear Physics of the U.S. Department of Energy
under Contract No. DE-AC03-76SF00098 and DE-FG05-90ER40592.
P. L. and X.-N. W. were also supported by the U.S. - Hungary
Science and Technology Joint Fund J. F. No. 93B/378.



\small

\begin{thebibliography}{99}
\baselineskip 9pt
\bibitem{KMR} J. Rafelski, Phys. Rep. {\bf 88} (1982) 331;
P. Koch, B. M\"uller, J. Rafelski, Phys. Rep. {\bf 142} (1986) 167;
QM'91 Conference, Nucl. Phys. {\bf A544}(1992) 1; QM'93 Conference, Nucl.
Phys. {\bf A566} (1994) 1.;
\bibitem{HIJING}X.-N.~Wang and M.~Gyulassy, Phys. Rev. D {\bf 44},
         3501 (1991); Comp. Phys. Commun. {\bf 83}, 307 (1994).
\bibitem{PCM}K.~Geiger and B.~M\"{u}ller, Nucl. Phys. {\bf B369}, 600 (1992);
        K.~Geiger, Phys. Rev.  D {\bf 47}, 133 (1993).
\bibitem{DTUNUC}H.~J.~Moehring and J.~Ranft, Z. Phys. C {\bf 52},
        643 (1991); P. Aurenche,{\em et al.}, Phys. Rev. D{\bf 45}, 92 (1992).
        P. Aurenche,{\em et al.}, Comp. Phys. Commun. {\bf 83}, 107 (1994).
\bibitem{JBAMKLL}J.P. Blaizot, A.H. Mueller, Nucl. Phys. {\bf B289},
847 (1987); K.{}~Kajantie, P.{}~V.{}~Landshoff and J.{}~Lindfors,
        Phys. Rev. Lett. {\bf 59}, 2517 (1987); K.{}~J.{}~Eskola,
        K.{}~Kajantie and J.{}~Lindfors, Nucl. Phys. {\bf B323}, 37 (1989);
        K.{}~J.{}~Eskola, K.{}~Kajantie and J.{}~Lindfors, Phys.
        Lett. B {\bf 214}, 613 (1989).
\bibitem{SHUR93}E. Shuryak, Phys. Rev. Lett. {\bf 68}, 3270 (1992);
   E. Shuryak and L. Xiong, Phys. Rev. Lett. {\bf 70} 2241 (1993);
   L. Xiong and E. Shuryak, Phys. Rev. C {\bf 49}, 2207 (1994).
\bibitem{BDMTW}T.~S.~Bir\'o, E.~van~Doorn, B.~M\"uller, M.~H.~Thoma,
        and X.-N.~Wang, Phys. Rev. C {\bf 48}, 1275 (1993).
\bibitem{BEW}S.~J.~Brodsky and H.~J.~Lu, Phys. Rev. Lett. {\bf 64},
        1342 (1990);
        K. J. Eskola, J. Qiu and X.-N. Wang, Phys. Rev. Lett.
        {\bf 72}, 36 (1994);
         X.-N. Wang and M. Gyulassy, Phys. Rev.  Lett. {\bf 68},
        1480 (1992).
\bibitem{KEXW94a}K. J. Eskola and X.-N. Wang, Phys. Rev. D {\bf 49},
        1284 (1994).
\bibitem{GWLPM}M.~Gyulassy and X.-N.~Wang, Nucl. Phys. B {\bf 420},
        583 (1994); X.-N.~Wang, M.~Gyulassy and M.~Pl\"umer, LBL-35980,
        hep-ph/9408344.
\bibitem{OCHARM}P.~L\'evai, B. ~M\"uller, X.-N. ~Wang, LBL-36594,
        hep-ph/9412352.
\bibitem{ALCOR} T.S. Bir\'o, P. L\'evai, J. Zim\'anyi, Phys. Lett. {\bf B347}
        6 (1995); T.S. Bir\'o, P. L\'evai, J. Zim\'anyi, hep-ph/9504203 and
        see this Proceedings.
\bibitem{BMW92}T. S. Bir\'o, B. M\"uller, and X.-N. Wang,
        Phys. Lett. {\bf B283}, 171 (1992).
\bibitem{KEMG}K. J. Eskola and M. Gyulassy, Phys. C {\bf 47}, 2329 (1993).
\bibitem{MSM86}T. Matsui, B. Svetitsky, and L. McLerran, Phys. Rev. D
        {\bf 34}, 783 (1986).
\bibitem{VISC} P. Danielewicz and M. Gyulassy, Phys. Rev. D {\bf 31},
        53 (1985); A. Hosoya and K. Kajantie, Nucl. Phys. {\bf B250},
        666 (1985); S. Gavin, Nucl. Phys. {\bf A435}, 826 (1985).
\bibitem{BJOR}J. D. Bjorken, Phys. Rev. D {\bf 27}, 140 (1983).
\bibitem{BP90}E. Braaten and R. D. Pisarski, Nucl. Phys. {\bf B337},
        569 (1990).
\bibitem{TBBM93}T. S. Bir\'o and B. M\"uller, Nucl. Phys. {\bf A561},
        477 (1993).
\bibitem{GUNION}J.~F.~Gunion and G.~Bertsch, Phys. Rev. D {\bf 25},
        746 (1982).
\end{thebibliography}
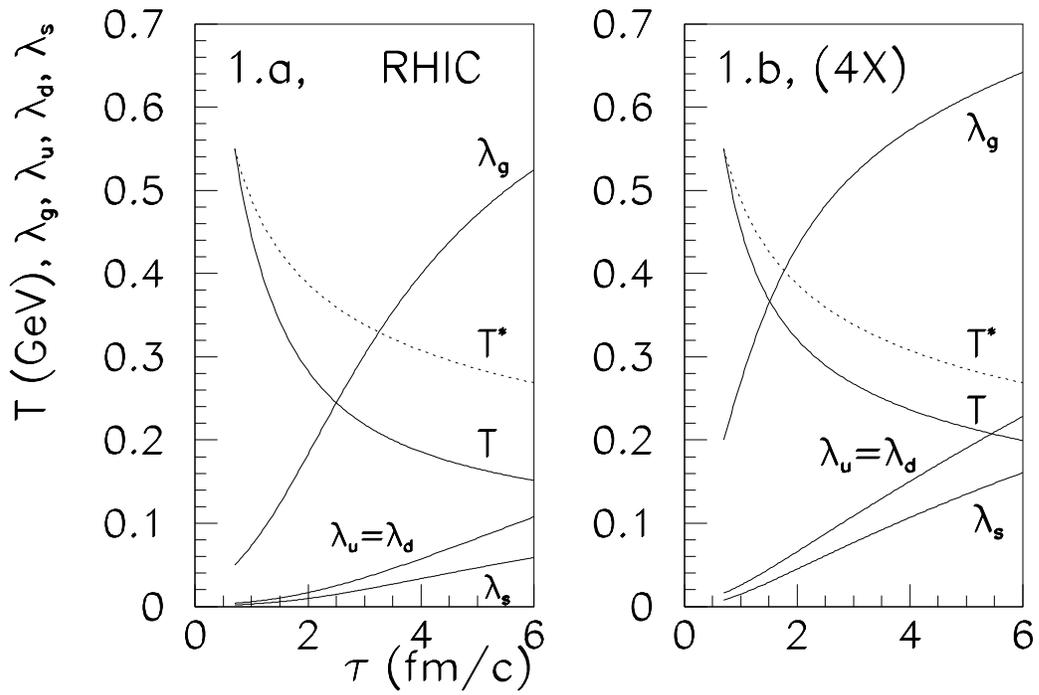
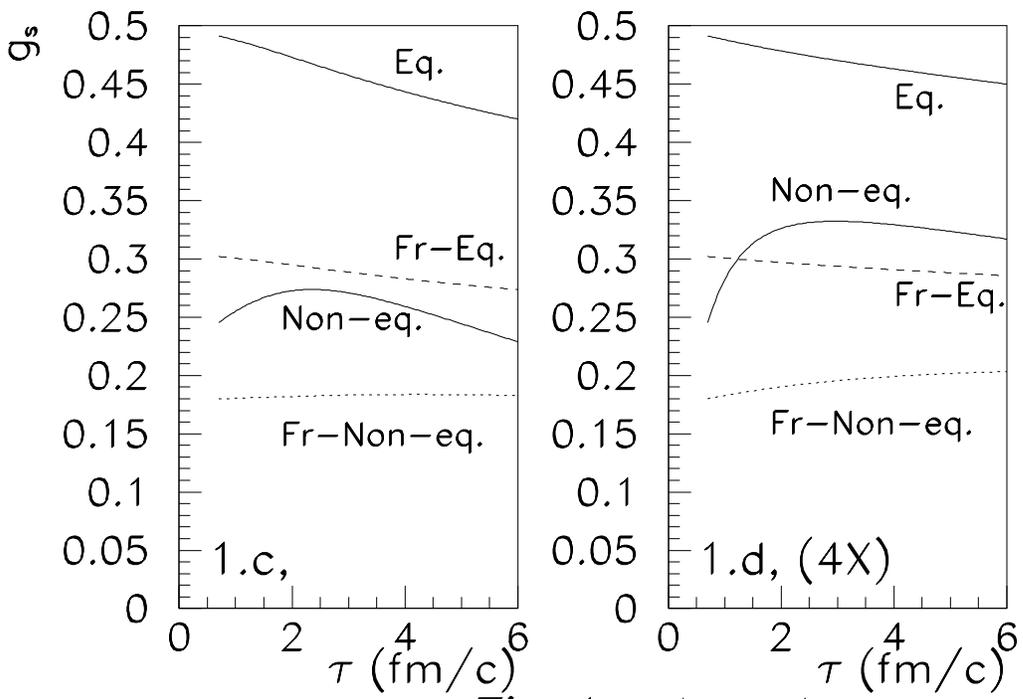
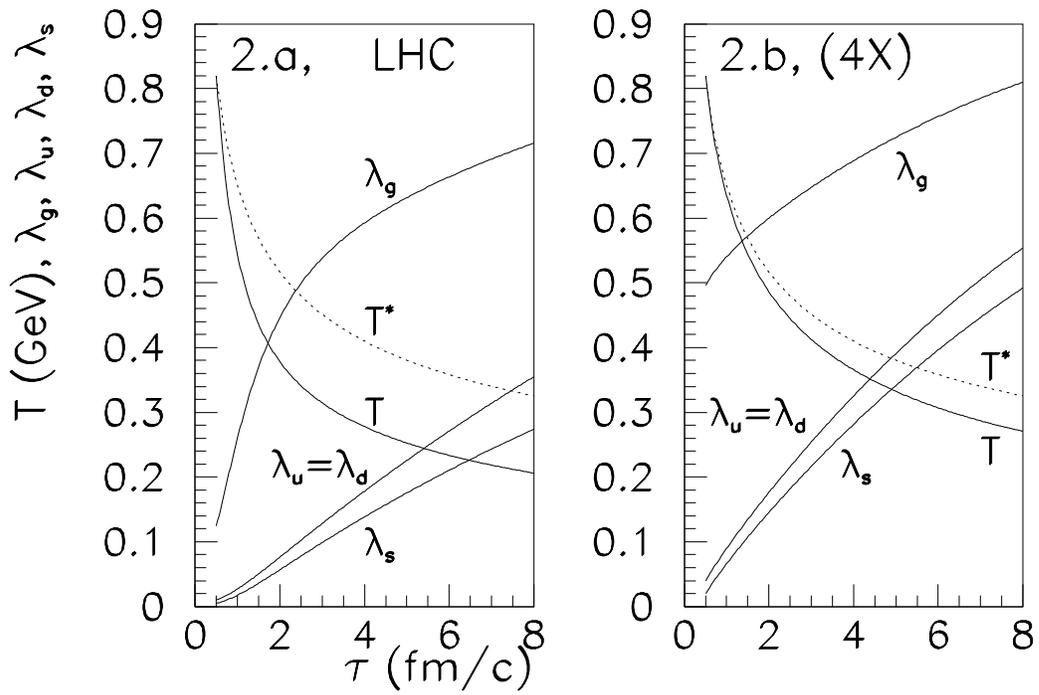
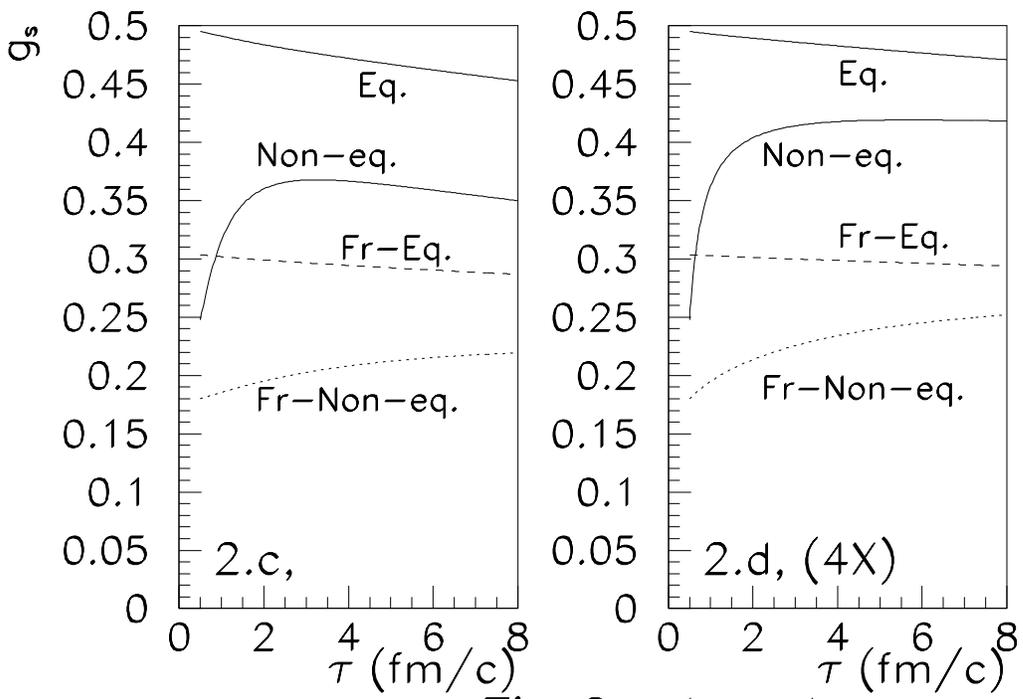
\end{document}